\documentclass[a4paper]{article}
\usepackage{xspace}
\usepackage{tikz}
\usepackage{xcolor}
\usepackage{dirtytalk}
\usepackage{fancyvrb}
\usepackage{hyperref}
\usepackage{url}
\usepackage[most,breakable]{tcolorbox}
\usepackage[utf8]{inputenc}
\usepackage{cprotect}

\definecolor{changes}{RGB}{255,255,255}
\definecolor{instance}{RGB}{250,220,220}
\definecolor{reduction}{RGB}{250,250,220}
\definecolor{certification}{RGB}{220,250,220}
\definecolor{correctness}{RGB}{150,250,150}

\newtheorem{definition}{Definition}
\newtheorem{example}{Example}

\newcommand{\ursa}{{\sc ursa}\xspace}
\newcommand{\sat}{{\sc sat}\xspace}
\newcommand{\dpll}{{\sc dpll}\xspace}
\newcommand{\np}{{\sc np}\xspace}
\newcommand{\p}{{\sc p}\xspace}
\newcommand{\cnf}{{\sc cnf}\xspace}

\title{A SAT-based Approach for Specification, Analysis, and 
Justification of Reductions between NP-complete Problems}
\author{Predrag Jani\v{c}i\'c \\
Faculty of Mathematics, University of Belgrade, \\ 
Studentski trg 16, 11000 Belgrade, Serbia
}
\date{}

\begin{document}

\maketitle

\begin{abstract}
We propose a novel framework for developing, analyzing, and validating 
reductions between \np-complete problems. Powered by the \sat-based 
constraint solver \ursa, our methodology introduces several distinct 
features that set it apart from other related approaches. The proposed 
workflow effectively bridges the crucial gap between informal, high-level 
reduction descriptions and formalized mathematical proofs. By supplementing 
rather than replacing human intuition, this interactive methodology serves 
as an aid for exploring relationships between \np-complete problems.
\end{abstract}

%\keywords{complexity classes, \np-completeness, \sat problem}
%\subjclass{F.2, F.1.3}

% ***************************************************************************
% ***************************************************************************
\section{Introduction}
\label{sec:introduction}
% ***************************************************************************
% ***************************************************************************

\np-completeness stands as a foundational concept in theoretical computer 
science and computational complexity theory \cite{Papadimitriou,ComplexityModern}. 
On a practical level, identifying a problem as \np-complete indicates that finding 
a polynomial-time algorithm for its exact solution is highly improbable (unless 
\p = \np). This understanding typically redirects efforts toward approximation 
algorithms or alternative real-world problem-solving strategies. Since the 
propositional satisfiability problem (\sat) was initially proven to be \np-complete 
\cite{Cook71,Levin}, the standard method for establishing the \np-completeness of 
a new problem involves two key steps: proving it belongs to \np, and demonstrating 
a polynomial-time reduction from a known \np-complete problem. While this 
methodology has been successfully applied to hundreds of problems 
\cite{garey-johnson,AlgorithmDesignKleinberg,Ruiz-VanoyeOPDSHRF11}, sharing some 
ideas and techniques, each proof 
typically demands a new non-trivial construction. Reductions are also broadly 
useful for determining membership in other complexity classes, establishing
decidability, and translating problems to leverage existing solvers \cite{Kneisel2024}.

Concepts in \np-completeness frequently lead to misconceptions and errors 
in computer science literature \cite{Mann17}.
Common pitfalls in proving that a problem is \np-complete include reducing 
in the wrong direction, failing to establish bidirectionality, or neglecting 
to prove that a problem is in \np. 
These challenges also create significant hurdles in computer science 
education; manual development and verification of reductions is often tedious, 
unreliable, and complicated by subtle logical gaps \cite{ZhangHD22,CreusFG14}.

There are very few dedicated toolkits or specification languages designed to 
develop, analyse, or verify \np-completeness reductions \cite{CreusFG14,ZhangHD22,GrangeVVZ24}. 
Existing systems are primarily utilized as educational prototypes, with 
no new \np-completeness results confirmed through their use. There are 
developments regarding complexity theory and the notion of \np-completeness 
within interactive theorem provers, offering the highest level of confidence 
\cite{CoqGaherK21,RassJHH24,LeanSimas}. Although some errors in traditional 
pen-and-paper proofs of \np-completeness were identified using interactive 
provers \cite{KreuzerN23,CoqKatz}, their usage remains largely confined to 
the interactive theorem-proving community. Consequently, the vast majority 
of published \np-completeness proofs still rely on informal descriptions of 
reductions, expressed in natural language that lacks strictness necessary to 
replicate them.

This paper introduces a new, \sat-based approach aimed at supporting the 
process of working with reductions between problems and proving \np-complete\-ness. 
While automation of verification of reductions is not supported in the above tools, 
the proposed approach easily enables the automated justification 
of reduction correctness for bounded input sizes. 
The proposed approach is primarily a systematic methodology that 
uses the existing constraint-solving tool \ursa. While other constraint-solving 
tools could be used in a similar manner, \ursa offers some specifics: it uses 
a C-like imperative-declarative language that enables concise specifications. 
While retaining the benefits of a declarative nature, the \ursa language 
enables the use of working C-code verbatim or almost verbatim, and there 
is no need to learn another, custom language. Due to restrictions in forming 
loops in \ursa, it is typically trivial to establish the time complexity of 
\ursa procedures, which is an important step in proving \np-completeness. 

This paper demonstrates the proposed approach through concrete examples, 
showing how to rigorously represent problem instances and certificates, 
solve \np-complete problems, represent reductions, and justify their
validity. While this size-bounded confirmation does not constitute a 
full correctness proof, it substantially bolsters confidence in the 
reduction. This creates a practical workflow bridging the gap between 
unverified manual proofs and strict foundational formalizations, 
supplementing human effort in designing reductions rather than replacing 
it.\footnote{In this work we assume that reductions are designed 
(and implemented) by a human, although there are approaches for finding 
reductions automatically \cite{CrouchIM10,JordanK13}.}

In the following text, we will consider several \np-problems from  
Karp's seminal list of 21 \np-complete problems \cite{Karp72}: \sat, 
3\sat, 3-colourability, $k$-clique, and $l$-vertex cover. Notably, 
\sat serves a dual purpose in this framework: \textit{(i)} it is 
analysed and used just as other \np-complete problems; \textit{(ii)} 
it acts as the underlying computational engine, as \ursa transforms 
problems into \sat instances to be resolved by underlying \sat solvers.

\paragraph{Overview of the paper.}
The remainder of the paper is organised as follows. In Section 
\ref{sec:background} we give a brief background on \np-completeness, the 
\sat problem, and the \ursa system. In Section \ref{sec:solving} we describe 
how to solve instances of an \np-complete problem using \ursa, while in 
Section \ref{sec:reductions} we describe how to represent reductions between 
\np-complete problems using \ursa. In Section \ref{sec:verification} we show 
how the correctness or soundness of reductions can be verified, and in Section 
\ref{sec:Workflow} we discuss the typical workflow. In Section \ref{sec:related} 
we discuss related work and in Section \ref{sec:conclusions} we draw the final 
conclusions.

% ***************************************************************************
% ***************************************************************************
\section{Background}
\label{sec:background}
% ***************************************************************************
% ***************************************************************************

\subsection{\np-completeness}

A decision problem is a question with a {\em yes} or {\em no} (or $1$ or 
$0$) answer 
for any given input. The class of problems known as \np (nondeterministic 
polynomial time) is the set of decision problems that a nondeterministic 
Turing machine can solve in polynomial time (in the size of the input). 
If a potential solution (a certificate, a witness) is given for an instance 
of an \np problem, it can be checked whether it is correct by a deterministic 
Turing machine in polynomial time  (in the size of the input). Somewhat 
more formally, these notions can be defined as follows \cite{ComplexityModern}.

\begin{definition}
A {\em decision problem} is a problem of computing a function 
$f : \{ 0, 1 \}^* \rightarrow \{ 0, 1 \}$. We identify such a function $f$ 
with the subset $L_f = \{ x \; | \; f (x) = 1 \}$ of $\{ 0, 1 \}^*$
and call such sets {\em languages} or {\em decision problems}. 
We identify the problem of, given $x$, computing $f(x)$ with the 
problem of deciding whether $x \in L_f$. 
\end{definition}

\begin{definition}
A language $L \subseteq \{ 0, 1 \}^*$ is in \np if there exists 
a polynomial $p : {\bf N} \rightarrow {\bf N}$ and a deterministic polynomial-time 
Turing machine $M$ such that for every $x \in \{ 0, 1 \}^*$ it holds:
$x \in L$ iff $\exists c_x \in \{0, 1\}^{p(|x|)}$ such that 
$M (x, c_x) = 1$. If $x \in L$ and $c_x \in \{0, 1\}^{p(|x|)}$
satisfy $M (x, c_x) = 1$ then we call $c_x$ a certificate for $x$ 
(with respect to the language $L$ and machine $M$).
\end{definition}

Note that checking a certificate is completely different from 
searching for it. If $x$ is an instance of the problem $X$, and 
if $c_x$ is a correct certificate for it, we will denote it by 
$Cert_X(x,c_x)$.

\begin{definition}
A language $L \subseteq \{ 0, 1 \}^*$ is polynomial-time reducible to 
a language $L' \subseteq \{ 0, 1 \}^*$, denoted by $L \leq_p L'$, if 
there is a polynomial-time computable function 
${\mathcal R} : \{ 0, 1 \}^* \rightarrow \{ 0, 1 \}^*$ such that for 
every $x \in \{ 0, 1 \}^*$,
$x \in L$ if and only if ${\mathcal R}(x) \in L'$.

We say that $L'$ is \np-hard if $L \leq_p L'$ for every $L \in$ \np. 
We say that $L'$ is \np-complete if $L'$ is \np-hard and $L' \in$ \np.
\end{definition}

In terms of decision problems, a reduction from $X$ to $Y$ is a function 
${\mathcal R}$ that transforms a valid input of $X$ into a valid input
of $Y$ and that preserves the answer: $x$ leads to answer {\em yes}
(for $X$) iff ${\mathcal R}(x)$ leads to answer {\em yes} (for $Y$).
A decision problem $Y$ is called \np-{\em hard} if any problem $X$ 
from \np can be reduced to it in polynomial time, denoted $X \leq_p Y$. 
Therefore, if one can solve an \np-hard problem $Y$ in polynomial 
time, they can also solve any other problem from the \np class in 
polynomial time. 
A problem is \np-{\em complete} if it is both in \np and is \np-hard. 
The propositional satisfiability problem (\sat) was the first problem proven to be 
\np-complete, and it remains a cornerstone of computational complexity theory 
\cite{Cook71,Levin}. To prove a new problem $Y$ is \np-complete, instead of
following the formal definition, it is sufficient to prove two things 
\cite{Papadimitriou,ComplexityModern}:

\begin{enumerate} 
\item $Y$ is in \np; 
\item for some \np-complete problem $X$ it holds $X \leq_p Y$.
\end{enumerate} 

\noindent
To show $X \leq_p Y$, it is sufficient, for an arbitrary instance $x$ of $X$,  
to construct, in polynomial time, an instance $y = \mathcal{R}(x)$ of $Y$
($\mathcal{R}$ is a reduction function), 
such that $x$ has the answer \say{yes} iff $y$ has the answer \say{yes}. In 
other words, the following has to be proved: 
$$\forall x \in X \;\; ((\exists c_x \; Cert_X(x,c_x)) \Leftrightarrow 
(\exists c_y \; Cert_Y(\mathcal{R}(x),c_y)))$$
In concrete proofs, however, instead of the {\em existence} of a certificate for 
$\mathcal{R}(x)$, a potential solution $c_y$ obtained by a translation 
closely related to $\mathcal{R}$ (hence we will also denote it by 
$\mathcal{R}$) is often considered, and it is proved:
$$\forall x \in X  \;\; \forall c_x \; (Cert_X(x,c_x) \Leftrightarrow Cert_Y(\mathcal{R}(x),\mathcal{R}(c_x)))$$

\noindent
The above reduction correctness condition has two directions, called 
{\em completeness}:
$$\forall x \in X  \;\; ((\exists c_x \; Cert_X(x,c_x)) \Rightarrow 
(\exists c_y \; Cert_Y(\mathcal{R}(x),c_y)))$$
\noindent or
$$Cert_X(x,c_x) \Rightarrow Cert_Y(\mathcal{R}(x),\mathcal{R}(c_x))$$
and {\em soundness}:
$$ \forall x \in X \;\; 
((\exists c_y  \; Cert_Y(\mathcal{R}(x),c_y)) \Rightarrow 
(\exists c_x  \; Cert_X(x,c_x)))$$
\noindent
or 
$$Cert_Y(\mathcal{R}(x),\mathcal{R}(c_x)) \Rightarrow Cert_X(x,c_x)$$

\begin{example}[Clique to Vertex Cover reduction]
\label{ex:clique_vertex-cover}
A {\em clique} of a graph $G$ is a subset $S$ of vertices such that it 
forms a complete subgraph in $G$. A {\em $k$-clique problem} is the decision 
problem of determining whether a graph $G$ has a clique with at least $k$ 
vertices (where $G$ and $k$ are inputs).\footnote{Note that 
{\em $k$-clique problem} for a fixed $k$ has polynomial time solutions.}

A {\em vertex cover} of a graph $G$ is a subset $S$ of vertices such that 
every edge has at least one end in $S$. An {\em $l$-vertex cover problem} 
is the decision problem of determining whether a graph $G$ has a vertex 
cover with at most $l$ vertices (where $G$ and $l$ are inputs). 

Let us prove that the {\em $l$-vertex cover problem} is \np-complete assuming 
that the {\em $k$-clique problem} is \np-complete.

\begin{enumerate} 
\item Given a potential solution, a set $S$, we can check in polynomial time 
all the edges and confirm if $S$ is indeed an $l$-vertex cover. Hence, 
the {\em $l$-vertex cover problem} is in \np.

\item Now, let us show $k$-clique $\leq_p$ $l$-vertex cover. Let $y$ be an instance 
of the $k$-clique problem: \say{does $G$ contain a clique of size at least $k$?},
where $G(V,E)$ is a graph with $n$ vertices. Let $G'$ be the complement graph 
for $G$: the sets of the vertices of $G$ and $G'$ are the same, and in $G'$ 
there is an edge between two vertices iff there is no edge between these two 
vertices in $G$. Obviously, $G'$ can be constructed in polynomial time. Now 
it holds: $G$ has a clique of size at least $k$ iff $G'$ has a vertex 
cover of size at most $l$, where  $l = n-k$. More concretely, $G$ has 
a clique $S$ of size at least $k$, iff $G'$ has a vertex cover 
$V \setminus S$, which is of size at most $n-k$. Let us prove that.

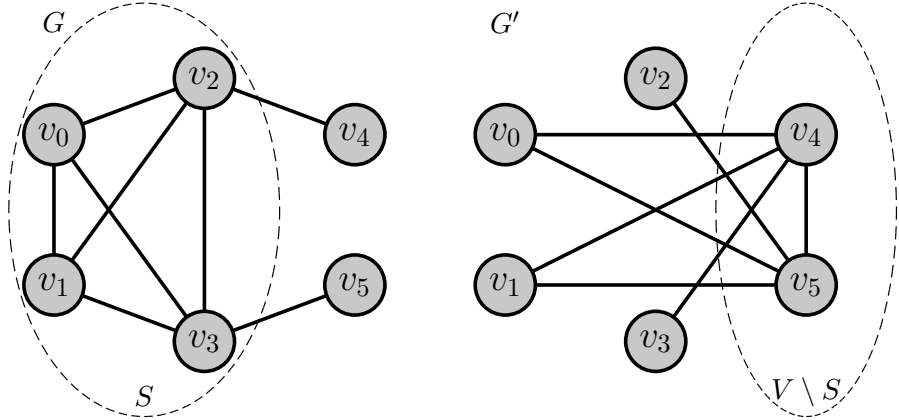
\begin{center}
\input{clique2vertexCover.tkz}
\end{center}

\begin{description}
\item[$\Rightarrow$]
Let us suppose that $G$ has a $k$-clique, made of a set of nodes $S$. None of 
the edges between these nodes belongs to $G'$. Let us prove that 
$V \setminus S$ indeed forms a vertex cover (of size at most $n-k$), i.e., 
let us prove that for an arbitrary edge $e$ from $G'$ at least one of its 
endpoints belongs to $V \setminus S$. Since $e$ belongs to $G'$, it does not 
belong to $G$. Therefore, at least one of its endpoints does not belong to $S$ 
(since $S$ forms a clique). Further, at least one of the endpoints of $e$ belongs 
to $V \setminus S$.

\item[$\Leftarrow$]
Let us suppose that $G'$ has a vertex cover $V \setminus S$ of size at most 
$n-k$. Let us prove that $S$ forms a $k$-clique in $G$. For an arbitrary edge 
from $G'$, one or both of its endpoints belong to $V \setminus S$. Thus, if neither 
of the two vertices of an edge belongs to $V \setminus S$ (i.e., if they belong to
$S$), then that edge does not belong to $G'$, so it belongs to $G$. In other 
words, if two nodes of $G$ belong to $S$, then the edge connecting them must 
be in $G$. 
\end{description}
\end{enumerate} 

If $x$ is an instance of the $k$-clique problem \say{does $G$ contain a 
clique of size at least $k$?}, and $y$ is a corresponding instance of the 
$l$-vertex cover problem \say{does $G'$ contain a cover of size at most 
$n-k$?}, using the above notation, $\mathcal{R}(x) = y$, 
$\mathcal{R}(S) = V\setminus S$ and

$$Cert_{k-clique}(x,S) 
  \Leftrightarrow 
Cert_{l-vertex\; cover}(y,V \setminus S)
 \Leftrightarrow   
Cert_{l-vertex\; cover}(\mathcal{R}(x),\mathcal{R}(S))$$
\end{example}

\subsection{\sat Problem and \sat Solvers.}

\sat is the problem of deciding whether a given propositional formula in 
conjunctive normal form (\cnf) is satisfiable, i.e., if there is any 
assignment to variables such that all clauses are true. Most of the 
state-of-the-art complete \sat solvers are {\sc cdcl} (conflict-driven, 
clause-learning) based extensions of the Davis-Putnam-Logemann-Loveland 
algorithm (\dpll) \cite{dp,DLL62,HandbookOfSAT2009}. In recent decades, 
great advances have been made in \sat solving technology 
\cite{MMZZM01,minisat,ZhangQuest02,HandbookOfSAT2009}. These advances 
in \sat solving make it possible to decide satisfiability of some industrial 
\sat problems with tens of thousands of variables and millions of clauses.

\subsection{System \ursa}

The open-source constraint solving system 
\ursa\footnote{\url{https://github.com/janicicpredrag/ursa}}
\cite{janicic_ursa_2012} is based on reduction to the propositional 
satisfiability problem (\sat). In \ursa, the problem is specified in 
a language that is imperative and similar to C, but at the same time 
is declarative, since the user does not have to provide a solving 
mechanism for the given problem. \ursa supports two types of variables: 
natural numbers (with names beginning with \verb|n|, e.g., \verb|nX|) 
and Booleans (with names beginning with \verb|b|, e.g., \verb|bX|), 
with a wide range of C-like operators (arithmetic, relational, logical, 
and bitwise). Variables can have concrete (ground) values, or symbolic 
values in which case they are represented internally by vectors of 
propositional formulae. These vectors have a fixed, given size (the 
default is 8). There is support for procedures, and there are control-flow 
structures (in the style of C). Loops must have known, ground bounds, and 
\verb|break| cannot be used. Also, if-statements can be applied only on 
ground expressions, and arrays can be accessed only with ground indices. 
Although they are not available in C (it would be natural if they were), 
in \ursa there is the operator \verb|^^| for logical exclusive disjunction, 
and compound assignment operators \verb|&&=|, \verb_||=_, \verb|^^=| with 
semantics defined in the same spirit as, for instance, \verb|+=| from C. 
There is an \verb|ite| function (corresponding to the operator \verb|?:| 
from C), and it can have any argument either ground or symbolic. An \ursa 
specification is symbolically executed \cite{King76,PasareanuV09} and the 
asserted condition given by the user is translated to a propositional 
formula. Then it is transformed into \cnf and passed to one of the 
underlying \sat solvers. If this formula is satisfiable, the system can 
return one model or all its models and provides values of the variables 
initially undefined.

\begin{example}
\label{ex:trivial}
Consider a trivial problem: if $nv$ equals $nu+1$, find a value for $nu$ 
such that $nv$ equals $2$. If a value of $nu$ is given in advance, one 
could easily check whether it is a solution of the problem. Indeed, one 
would assign $nu+1$ to $nv$ and finally check whether $nv$ equals $2$. 
Such a test can be written in the form of an imperative C-like code (where 
{\tt assert(b)} checks whether {\tt b} is true) as follows:

\begin{tcolorbox}
{\scriptsize
\begin{verbatim}
nv = nu+1;
assert(nv==2);
\end{verbatim}
}
\end{tcolorbox}

\noindent
However, since \verb|nu| is not given, \ursa treats it as unknown, translates 
the condition \verb|nv==2| to \sat, solves it, and uses the model found to 
provide the sought value for \verb|nu|: 1.
\end{example}

\begin{example}
A linear congruential pseudorandom number generator is defined by 
a recurrence relation of the form:

$x_{n+1} \equiv (a x_n + c) \;\; (\mathrm{mod} \; m) \;\;\; (\mbox{for} \;\; n \geq 0)$

\noindent
where $x_0$ is the {\em seed} value ($0 \leq x_0 < m$). One example
of such relation is:

$x_{n+1} \equiv (1664525 x_n + 1013904223) \;\; (\mathrm{mod} \; 2^{32}) \;\;\; (\mbox{for} \;\; n \geq 0)$

\noindent
It is trivial to compute elements of this sequence if the seed value 
is given. The check that $x_{100}$ is indeed equal to the given value 
if the seed is equal to $nseed$ can be simply written in the form 
of an imperative C-like code as follows (assuming that numbers are 
represented by $32$ bits, and \verb|nseed| is a concrete value):

\begin{tcolorbox}
{\scriptsize
\begin{verbatim}
nx=nseed;
for(ni=1;ni<=100;ni++)
nx=nx*1664525+1013904223;
assert(nx==2365677197);
\end{verbatim}
}
\end{tcolorbox}

\noindent
However, it is non-trivial to compute $x_0$ given the value $x_{100}$. 
Still, the very same test shown above can serve as a specification of 
this problem, with \verb|nseed| kept undefined. The \ursa system, using 
the above specification with the representation length $32$, generates 
a formula with 108132 variables and 366532 clauses. The solution of the 
problem (a required value for \verb|nseed|: 2025), is found in less than 
0.5~s (including the generating and solving phase).
\end{example}

Note that the specifications given above also cover the information on 
what variables are unknown and have to be determined so that the constraints 
are satisfied---those are variables that appear within commands before 
they are defined. So, the above code is a full and precise specification 
of the problem, up to the interval of the variables (i.e., the length of
vectors of formulae).

The \ursa system has been used for solving problems from various domains
including cryptography \cite{Kircanski15}, geometric combinatorial problems
\cite{ConstructibilityClasses}, chess \cite{chess-ursa}, and automated 
theorem proving \cite{janicic_theorem_2022}.

% ***************************************************************************
% ***************************************************************************
\section{Solving \np-complete Problems using \ursa}
\label{sec:solving}
% ***************************************************************************
% ***************************************************************************

The imperative-declarative nature of the \ursa system allows for an 
elegant solution to \np-complete problem instances. Due to \ursa's 
declarative nature, implementing a problem solver requires only specifying 
a certificate checking procedure rather than complex search techniques. 
Furthermore, its imperative features enable the certificate check 
implementation to be very similar (or even identical) to one implemented 
in a language like C.

\begin{example}[$k$-clique solver]
\label{ex:k-clique}
In order to prove that the $k$-clique problem belongs to \np, one must 
demonstrate that it admits a polynomial-time verification of a certificate 
(in terms of the input size). 
The input graph is given by an adjacency matrix; therefore, along with 
the representation of $k$, the encoding of the input requires $O(|V|^2)$ 
Booleans or bits. 
A suitable certificate $S$ is a subset of 
the graph's nodes. The required check is to verify that $S$ forms a 
complete subgraph (a clique) and that its size is at least $k$. We can 
easily specify the certificate verification procedure in \ursa, as shown 
in Figure \ref{fig:k-clique}.

\begin{figure}
\begin{minipage}{0.54\textwidth}
\begin{tcolorbox}[left=0mm,right=0mm]
{\scriptsize
\begin{tcolorbox}[colback=instance,left=0mm,boxrule=0.1pt]
\begin{verbatim}
/* Specification of instance of k-clique */
nV = 6; 
nK_clique = 4; 

bE_clique[0][1] = true;
bE_clique[0][2] = true;
bE_clique[0][3] = true;
bE_clique[0][4] = false;
bE_clique[0][5] = false;
bE_clique[1][2] = true;
bE_clique[1][3] = true;
bE_clique[1][4] = false;
bE_clique[1][5] = false;
bE_clique[2][3] = true;
bE_clique[2][4] = true;
bE_clique[2][5] = false;
bE_clique[3][4] = false;
bE_clique[3][5] = true;
bE_clique[4][5] = false;
\end{verbatim}
\end{tcolorbox}

\begin{tcolorbox}[colback=changes,left=0mm,boxrule=0.1pt]
\begin{Verbatim}[formatcom=\color{black}]
/* Specification of certificate for k-clique */
bBelongsClique[0]=true;
bBelongsClique[1]=true;
bBelongsClique[2]=true;
bBelongsClique[3]=true;
bBelongsClique[4]=false;
bBelongsClique[5]=false;
\end{Verbatim}
\end{tcolorbox}

\begin{tcolorbox}[colback=certification,left=0mm,boxrule=0.1pt]
\begin{verbatim}
/* Certificate verification for k-clique */
nCount = 0;
bClique = true;
for(ni=0; ni<nV; ni++) {
  nCount += ite(bBelongsClique[ni], 1, 0);
  for(nj=ni+1; nj<nV; nj++) {
    bClique &&= ite(
      bBelongsClique[ni] && bBelongsClique[nj], 
      bE_clique[ni][nj], true);
  }
}
bClique &&= (nCount >= nK_clique);
\end{verbatim}
\end{tcolorbox}

\begin{tcolorbox}[colback=certification,frame hidden,left=0mm,boxrule=0.1pt]
\begin{Verbatim}
print bClique;
\end{Verbatim}
\end{tcolorbox}
}
\end{tcolorbox}
\end{minipage}
\begin{minipage}{0.54\textwidth}
\begin{tcolorbox}[left=0mm,right=0mm]
{\scriptsize
\begin{tcolorbox}[colback=instance,left=0mm,boxrule=0.1pt]
\begin{verbatim}
/* Specification of instance of k-clique */
nV = 6; 
nK_clique = 4; 

bE_clique[0][1] = true;
bE_clique[0][2] = true;
bE_clique[0][3] = true;
bE_clique[0][4] = false;
bE_clique[0][5] = false;
bE_clique[1][2] = true;
bE_clique[1][3] = true;
bE_clique[1][4] = false;
bE_clique[1][5] = false;
bE_clique[2][3] = true;
bE_clique[2][4] = true;
bE_clique[2][5] = false;
bE_clique[3][4] = false;
bE_clique[3][5] = true;
bE_clique[4][5] = false;
\end{verbatim}
\end{tcolorbox}

\begin{tcolorbox}[colback=changes,frame hidden,left=0mm,boxrule=0.1pt]
\begin{Verbatim}[formatcom=\color{changes}]
/*** Specification of certificate for k-clique ***/
bBelongsClique[0]=true;
bBelongsClique[1]=true;
bBelongsClique[2]=true;
bBelongsClique[3]=true;
bBelongsClique[4]=false;
bBelongsClique[5]=false;
\end{Verbatim}
\end{tcolorbox}

\begin{tcolorbox}[colback=certification,left=0mm,boxrule=0.1pt]
\begin{verbatim}
/* Certificate verification for k-clique */
nCount = 0;
bClique = true;
for(ni=0; ni<nV; ni++) {
  nCount += ite(bBelongsClique[ni], 1, 0);
  for(nj=ni+1; nj<nV; nj++) {
    bClique &&= ite(
      bBelongsClique[ni] && bBelongsClique[nj], 
      bE_clique[ni][nj], true);
  }
}
bClique &&= (nCount >= nK_clique);
\end{verbatim}
\end{tcolorbox}

\begin{tcolorbox}[colback=certification,frame hidden,left=0mm,boxrule=0.1pt]
\begin{Verbatim}
assert(bClique);
\end{Verbatim}
\end{tcolorbox}
}

\end{tcolorbox}
\end{minipage}
\caption{Code for solving $k$-clique problem: verifying that a
concrete certificate is correct (left) and finding a certificate
for given instance (right)}
\label{fig:k-clique}
\end{figure}

\noindent
In the shown example (left),\footnote{\ursa specifications are stored as 
plain text files, colour annotations given here are used for presentation
purposes only.}
the given graph has 6 nodes (denoted $0$, $1$, $2$, $3$, $4$, $5$), $k$ 
i.e., \verb|nK_clique| equals $4$ (the same as in Example 
\ref{ex:clique_vertex-cover}), the adjacency matrix (i.e., its proper 
upper part, since the graph is undirected) is given by the array 
\verb|bE_clique|, and the certificate is given by the array 
\verb|bBelongsClique| stating which nodes belong to the (potential) 
clique. The certificate verification, given at the end, clearly requires 
a running time quadratic in $|V|$; therefore it is polynomial in the input
size, and the $k$-clique problem indeed belongs to \np.

At the end of the shown code, the variable \verb|bClique| will have the 
value \verb|true|. All of the above is straightforward and looks, more or 
less, like a certificate verification code that one would write in C. But 
\ursa allows omitting assignments to elements of the certificate --- 
\verb|bBelongsClique| (Figure \ref{fig:k-clique}, right). Then, its 
elements will be treated as symbolic values, and if \verb|bClique| is 
asserted (at the end), \ursa will try to find values of symbolic variables 
such that \verb|bClique| is true. \ursa translates this task into a \sat 
instance, solves it, and provides the following result, i.e., a correct 
certificate/solution:

\begin{tcolorbox}[left=0mm]
{\scriptsize
\begin{verbatim}
bBelongsClique[0]=true;
bBelongsClique[1]=true;
bBelongsClique[2]=true;
bBelongsClique[3]=true;
bBelongsClique[4]=false;
bBelongsClique[5]=false;
\end{verbatim}
}
\end{tcolorbox}

The above confirms that there is a sought clique of size at least $4$, so 
for the given instance, the answer is \say{yes}. 
\end{example}

The argumentation for polynomial time complexity in the following 
examples is similar to the one above. For brevity, we will not 
detail it again.

\begin{example}[$l$-vertex cover solver]
\label{ex:l-vertexcover}
The \ursa code shown in Figure \ref{fig:l-cover} gives a solver for 
the $l$-vertex cover problem, with one concrete input instance given (the same 
as in Example \ref{ex:clique_vertex-cover}). A certificate is formed by 
elements of the array \verb|bBelongs_vertexCover| --- those that belong 
to the vertex cover.

\begin{figure}
\begin{tcolorbox}[left=0mm,right=0mm]
{\scriptsize
\begin{tcolorbox}[colback=instance,left=0mm,bottom=1mm,boxrule=0.1pt]
\begin{verbatim}
/*** Specification of instance of l-vertexCover ***/
nV = 6; 
nL_vertex_cover = 2; 

bE_vertexCover[0][1] = false;
bE_vertexCover[0][2] = false;
bE_vertexCover[0][3] = false;
bE_vertexCover[0][4] = true;
bE_vertexCover[0][5] = true;
bE_vertexCover[1][2] = false;
bE_vertexCover[1][3] = false;
bE_vertexCover[1][4] = true;
bE_vertexCover[1][5] = true;
bE_vertexCover[2][3] = false;
bE_vertexCover[2][4] = false;
bE_vertexCover[2][5] = true;
bE_vertexCover[3][4] = true;
bE_vertexCover[3][5] = false;
bE_vertexCover[4][5] = true;
\end{verbatim}
\end{tcolorbox}

\begin{tcolorbox}[colback=certification,left=0mm,bottom=1mm,boxrule=0.1pt]
\begin{verbatim}
/*** Certificate verification for l-vertexCover ***/
nCount = 0;
bVertexCover = true;
for(ni=0; ni<nV; ni++) {
  nCount += ite(bBelongs_vertexCover[ni], 1, 0);
  for(nj=ni+1; nj<nV; nj++) {
    bVertexCover &&= ite(bE_vertexCover[ni][nj], 
      bBelongs_vertexCover[ni] || bBelongs_vertexCover[nj], true);
  }
}
bVertexCover &&= (nCount <= nL_vertex_cover);

assert(bVertexCover);
\end{verbatim}
\end{tcolorbox}
}
\end{tcolorbox}
\caption{Code for solving $l$-vertex cover problem}
\label{fig:l-cover}
\end{figure}

For the given instance, \ursa returns the following solution:

\begin{tcolorbox}[left=0mm]
{\scriptsize
\begin{verbatim}
bBelongs_vertexCover[0]=false;
bBelongs_vertexCover[1]=false;
bBelongs_vertexCover[2]=false;
bBelongs_vertexCover[3]=false;
bBelongs_vertexCover[4]=true;
bBelongs_vertexCover[5]=true;
\end{verbatim}
}
\end{tcolorbox}
\end{example}

The above examples illustrate how an instance of an \np-complete problem
$X$ can be easily solved using \ursa, by giving only the certificate 
verification procedure. The general scheme consists of two modules:

\begin{center}
\begin{minipage}{0.6\textwidth}
\begin{tcolorbox}

\begin{tcolorbox}[colback=instance,left=0mm,boxrule=0.1pt]
Specification of input instance of $X$
\end{tcolorbox}

\begin{tcolorbox}[colback=certification,left=0mm,boxrule=0.1pt]
Certificate verification for $X$
\end{tcolorbox}

\end{tcolorbox}
\end{minipage}
\end{center}

The approach described above provides a method for solving instances of 
an \np-complete problem. Because the solving process utilizes the \ursa 
system and then proceeds via the \sat problem, it is unlikely to be as 
computationally efficient as an optimized, custom-designed solution. 
Nonetheless, this method remains valuable for smaller instances or as a 
rapid alternative when a specialized custom solver is not readily available.

% ***************************************************************************
% ***************************************************************************
\section{Reductions between \np-complete Problems using \ursa}
\label{sec:reductions}
% ***************************************************************************
% ***************************************************************************

Given the components described in Section \ref{sec:solving}, and the 
reduction components for pairs of problems, we can easily combine them 
to yield new solvers for \np-complete problems via reductions. 
A general scheme is as follows:

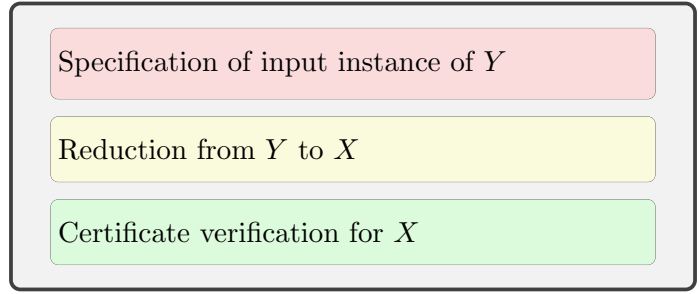
\begin{center}
\begin{minipage}{0.6\textwidth}
\begin{tcolorbox}

\begin{tcolorbox}[colback=instance,left=0mm,boxrule=0.1pt]
Specification of input instance of $X$
\end{tcolorbox}

\begin{tcolorbox}[colback=reduction,left=0mm,boxrule=0.1pt]
Reduction from $X$ to $Y$
\end{tcolorbox}

\begin{tcolorbox}[colback=certification,left=0mm,boxrule=0.1pt]
Certificate verification for $Y$
\end{tcolorbox}

\end{tcolorbox}
\end{minipage}
\end{center}

\begin{example}[$k$-clique to $l$-vertex cover reduction]
\label{ex:clique-cover-reduction}
Figure \ref{fig:clique-cover-reduction} shows how the reduction of 
the $k$-clique problem to the $l$-vertex cover problem, as described in Section 
\ref{sec:background}, can be implemented in \ursa.

The input instance of the $k$-clique problem is given by the size of the 
graph, the value $k$, and the proper upper part of the adjacency matrix 
(\verb|bE_clique|) (the same as in Example \ref{ex:k-clique}). This 
input instance is transformed into an input instance of the $l$-vertex 
cover problem, with the graph size $n$ the same as for the input instance, 
$l = n-k$, and the appropriate proper upper part of the adjacency matrix 
(\verb|bE_vertexCover|). The certificate verification code for the 
$l$-vertex problem is as in Example \ref{ex:l-vertexcover}.

\begin{figure}
\begin{tcolorbox}[left=0mm,right=0mm]
{\scriptsize

\begin{tcolorbox}[colback=instance,left=0mm,boxrule=0.1pt]
\begin{verbatim}
/*** Specification of instance of k-clique ***/
nV = 6; 
nK_clique = 4; 

bE_clique[0][1] = true;
bE_clique[0][2] = true;
bE_clique[0][3] = true;
bE_clique[0][4] = false;
bE_clique[0][5] = false;
bE_clique[1][2] = true;
bE_clique[1][3] = true;
bE_clique[1][4] = false;
bE_clique[1][5] = false;
bE_clique[2][3] = true;
bE_clique[2][4] = true;
bE_clique[2][5] = false;
bE_clique[3][4] = false;
bE_clique[3][5] = true;
bE_clique[4][5] = false;
\end{verbatim}
\end{tcolorbox}

\begin{tcolorbox}[colback=reduction,left=0mm,boxrule=0.1pt]
\begin{verbatim}
/*** Reduction from k-clique to l-vertexCover ***/
nL_vertexCover = nV - nK_clique;
for(ni=0; ni<nV; ni++) {
  for(nj=ni+1; nj<nV; nj++) {
    bE_vertexCover[ni][nj] = !bE_clique[ni][nj];
  }
}
\end{verbatim}
\end{tcolorbox}

\begin{tcolorbox}[colback=certification,left=0mm,boxrule=0.1pt]
\begin{verbatim}
/*** Certificate verification for l-vertexCover ***/
nCount = 0;
bVertexCover = true;
for(ni=0; ni<nV; ni++) {
  nCount += ite(bBelongs_vertexCover[ni], 1, 0);
  for(nj=ni+1; nj<nV; nj++) {
    bVertexCover &&= ite(bE_vertexCover[ni][nj], 
      bBelongs_vertexCover[ni] || bBelongs_vertexCover[nj], true);
  }
}
bVertexCover &&= (nCount <= nL_vertexCover);

assert(bVertexCover);
\end{verbatim}
\end{tcolorbox}
}
\end{tcolorbox}
\caption{Code for $k$-clique to $l$-vertex cover reduction}
\label{fig:clique-cover-reduction}
\end{figure}

For the given instance, \ursa returns a positive answer and (again) 
the following solution (for the target, $l$-vertex cover problem):

\begin{tcolorbox}[left=0mm]
{\scriptsize
\begin{verbatim}
bBelongs_vertexCover[0]=false;
bBelongs_vertexCover[1]=false;
bBelongs_vertexCover[2]=false;
bBelongs_vertexCover[3]=false;
bBelongs_vertexCover[4]=true;
bBelongs_vertexCover[5]=true;
\end{verbatim}
}
\end{tcolorbox}

Notice that the certificate verification procedure confirms that the 
$l$-vertex cover problem is in \np, while the reduction procedure 
establishes that it is \np-hard; namely, both procedures clearly retain a 
polynomial running time relative to the input size. Therefore, if the 
$k$-clique problem is \np-complete, so is the $l$-vertex cover problem.
\end{example}

\begin{example}[$l$-vertex cover to $k$-clique reduction]
In the same style as in the previous example, we can easily implement a 
reduction in the opposite direction: from the $l$-vertex cover problem to the $k$-clique
problem. The \ursa code, given in Figure \ref{fig:cover-clique-reduction},
is again self-explanatory. 

\begin{figure}
\begin{tcolorbox}[left=0mm,right=0mm]
{\scriptsize

\begin{tcolorbox}[colback=instance,left=0mm,boxrule=0.1pt]
\begin{verbatim}
/*** Specification of instance of l-vertexCover ***/
nV = 6; 
nL_vertexCover = 2; 

bE_vertexCover[0][1] = false;
bE_vertexCover[0][2] = false;
bE_vertexCover[0][3] = false;
bE_vertexCover[0][4] = true;
bE_vertexCover[0][5] = true;
bE_vertexCover[1][2] = false;
bE_vertexCover[1][3] = false;
bE_vertexCover[1][4] = true;
bE_vertexCover[1][5] = true;
bE_vertexCover[2][3] = false;
bE_vertexCover[2][4] = false;
bE_vertexCover[2][5] = true;
bE_vertexCover[3][4] = true;
bE_vertexCover[3][5] = false;
bE_vertexCover[4][5] = true;
\end{verbatim}
\end{tcolorbox}

\begin{tcolorbox}[colback=reduction,left=0mm,boxrule=0.1pt]
\begin{verbatim}
/*** Reduction from l-vertexCover to k-clique ***/
nK_clique = nV - nL_vertexCover; 
for(ni=0; ni<nV; ni++) {
  for(nj=ni+1; nj<nV; nj++) {
    bE_clique[ni][nj] = !bE_vertexCover[ni][nj];
  }
}
\end{verbatim}
\end{tcolorbox}

\begin{tcolorbox}[colback=certification,left=0mm,boxrule=0.1pt]
\begin{verbatim}
/*** Certificate verification for k-clique ***/
nCount = 0;
bClique = true;
for(ni=0; ni<nV; ni++) {
  nCount += ite(bBelongs[ni], 1, 0);
  for(nj=ni+1; nj<nV; nj++) {
    bClique &&= ite(bBelongs[ni] && bBelongs[nj], bE_clique[ni][nj], true);
  }
}
bClique &&= (nCount >= nK_clique);

assert(bClique);
\end{verbatim}
\end{tcolorbox}
}
\end{tcolorbox}
\caption{Code for $l$-vertex cover to $k$-clique reduction}
\label{fig:cover-clique-reduction}
\end{figure}

For the given instance, \ursa returns the positive answer and (again) 
the following solution (for the target problem):

\begin{tcolorbox}[left=0mm]
{\scriptsize
\begin{verbatim}
bBelongs[0]=true;
bBelongs[1]=true;
bBelongs[2]=true;
bBelongs[3]=true;
bBelongs[4]=false;
bBelongs[5]=false;
\end{verbatim}
}
\end{tcolorbox}

\end{example}

The above approach allows us to develop another new family of \np-complete 
problem solvers that rely on reductions (notice that a solver for $X$ is 
obtained as a by-product of proving that $Y$ is \np-complete). Because these
solvers involve \ursa processing, reduction and, finally, \sat solving---they 
are unlikely to surpass the efficiency of optimized, dedicated solvers. 
Conversely, the primary advantage of this solving paradigm lies in the rapid 
prototyping of reductions and the provision for their justification. 
Specifically, if a reduction from problem $X$ to $Y$ is developed, one can 
test it using a set of input instances, by confirming it produces the same 
\say{yes}/\say{no} output as a solver for $X$ directly encoded in \ursa 
(without the need for a dedicated solver for either $Y$ or $X$).

% ***************************************************************************
% ***************************************************************************
\section{Verification of Reductions between \np-complete Problems using \ursa}
\label{sec:verification}
% ***************************************************************************
% ***************************************************************************

The correctness of a reduction from problem $X$ to $Y$, when specified in 
the \ursa language, can, as mentioned previously, be readily tested for 
individual instances. However, it would be significantly more valuable if 
the reduction could be justified automatically, at least to a certain 
extent. We will demonstrate that this is indeed possible. The overall 
scheme is as follows:

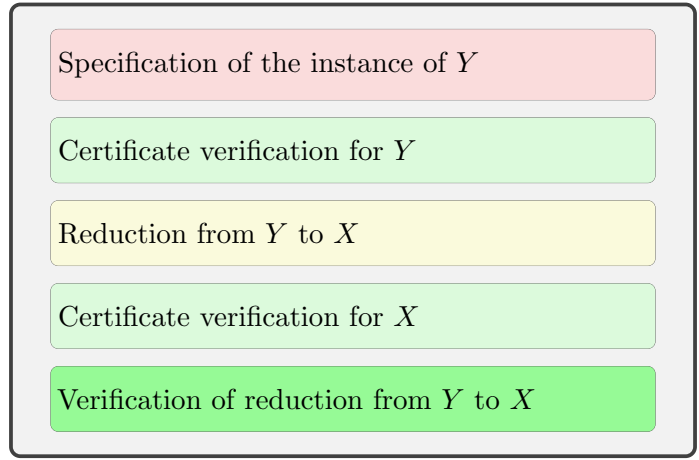
\begin{center}
\begin{minipage}{0.65\textwidth}
\begin{tcolorbox}

\begin{tcolorbox}[colback=instance,left=0mm,boxrule=0.1pt]
Specification of the instance of $X$
\end{tcolorbox}

\begin{tcolorbox}[colback=certification,left=0mm,boxrule=0.1pt]
Certificate verification for $X$
\end{tcolorbox}

\begin{tcolorbox}[colback=reduction,left=0mm,boxrule=0.1pt]
Reduction from $X$ to $Y$
\end{tcolorbox}

\begin{tcolorbox}[colback=certification,left=0mm,boxrule=0.1pt]
Certificate verification for $Y$
\end{tcolorbox}

\begin{tcolorbox}[colback=correctness,left=0mm,boxrule=0.1pt]
Verification of reduction from $X$ to $Y$
\end{tcolorbox}

\end{tcolorbox}
\end{minipage}
\end{center}

\begin{example}[Verification of $k$-clique to $l$-vertex cover reduction]
\label{ex:clique_to_vertexcover_verification}
Let us illustrate the verification mechanism on the reduction of the 
$k$-clique to the $l$-vertex cover problem, as given in Example 
\ref{ex:clique-cover-reduction}. Consider the \ursa code in Figure 
\ref{fig:clique-cover-reduction-verification}, with most of the fragments 
already presented.

\begin{figure}	
\begin{center}
\begin{tcolorbox}[left=0mm,right=0mm]
{\scriptsize

\begin{tcolorbox}[colback=instance,left=0mm,boxrule=0.1pt]
\begin{verbatim}
/*** Specification of the size of instance of k-clique ***/
nV = 6; 
nK_clique = 4; 
\end{verbatim}
\end{tcolorbox}

\begin{tcolorbox}[colback=certification,left=0mm,boxrule=0.1pt]
\begin{verbatim}
/*** Certificate verification for k-clique ***/
nCount = 0;
bClique = true;
for(ni=0; ni<nV; ni++) {
  nCount += ite(bBelongsClique[ni], 1, 0);
  for(nj=ni+1; nj<nV; nj++) {
    bClique &&= ite(bBelongsClique[ni] && bBelongsClique[nj], bE_clique[ni][nj], true);
  }
}
bClique &&= (nCount >= nK_clique);
\end{verbatim}
\end{tcolorbox}

\begin{tcolorbox}[breakable,colback=reduction,left=0mm,boxrule=0.1pt]
\begin{verbatim}
/*** Reduction from k-clique to l-vertexCover ***/
nL_vertexCover = nV - nK_clique;
for(ni=0; ni<nV; ni++) {
  for(nj=ni+1; nj<nV; nj++) {
    bE_vertexCover[ni][nj] = !bE_clique[ni][nj];
  }
}

/*** Link between certificate for k-clique and certificate for l-vertexCover ***/
bCertificateReduction = true;
for(ni=0; ni<nV; ni++) {
  bCertificateReduction &&= (bBelongsVertexCover[ni] ^^ bBelongsClique[ni]);
}
\end{verbatim}
\end{tcolorbox}

\begin{tcolorbox}[colback=certification,left=0mm,boxrule=0.1pt]
\begin{verbatim}
/*** Certificate verification for l-vertexCover ***/
nCount = 0;
bVertexCover = true;
for(ni=0; ni<nV; ni++) {
  nCount += ite(bBelongsVertexCover[ni], 1, 0);
  for(nj=ni+1; nj<nV; nj++) {
    bVertexCover &&= ite(bE_vertexCover[ni][nj], 
                         bBelongsVertexCover[ni] || bBelongsVertexCover[nj], 
                         true);
  }
}

bVertexCover &&= (nCount <= nL_vertexCover);
\end{verbatim}
\end{tcolorbox}

\begin{tcolorbox}[colback=correctness,left=0mm,boxrule=0.1pt]
\begin{verbatim}
assert((bClique ^^ bVertexCover) && bCertificateReduction);
\end{verbatim}
\end{tcolorbox}
}
\end{tcolorbox}
\end{center}
\caption{Code for verification of $k$-clique to $l$-vertex cover reduction }
\label{fig:clique-cover-reduction-verification}
\end{figure}

The final value of \verb|bClique| evaluates to a propositional formula that is 
satisfiable iff, for the given (symbolic) input instance, the answer is \say{yes}. 
Conversely, the final value of \verb|bVertexCover| is a propositional 
formula that is satisfiable iff, for the reduced input instance, the 
answer is \say{yes}. Notice that in the above \ursa code, only the size 
of the input instance is provided. Consequently, the elements of the 
adjacency matrix \verb|bE_clique| are treated as symbolic values, meaning 
that \verb|bClique| and \verb|bVertexCover| collectively encompass all 
possible input graphs of the given size. For checking correctness, one 
might be tempted to check {\em incorrectness}: to assert that 
\verb|bClique| and \verb|bVertexCover| are different 
(\verb|assert(bClique ^^ bVertexCover)|), hoping that the resulting 
unsatisfiability (\say{no solution}), i.e., absence of counterexamples, 
would ensure the correctness of the reduction. However, this is wrong: 
the formulae \verb|bClique| and \verb|bVertexCover| are not necessarily 
logically equivalent; rather, they are {\em equisa\-tisfi\-able}---one is 
satisfiable if and only if the other is satisfiable. Therefore, the 
assertion \verb|assert(bClique ^^ bVertexCover)| could succeed and yield 
a correct certificate for one problem and a wrong certificate for another, 
which proves nothing. In this case, one must also ensure the relationship 
between the certificates for the two problems and add that condition to 
the assertion: 
\verb|assert((bClique ^^ bVertexCover) && bCertificateReduction)|.
This assertion has no solutions which confirms correctness for $k=4$, 
for \emph{any} graph with $6$ vertices. By changing the constants 
\verb|nV| and \verb|nK_clique|, we can check correctness for any input 
size. For instance, for $k=20$ and graphs with $50$ vertices, for the 
given assertion \ursa generates a formula with $6430$ variables and 
$21811$ clauses. An underlying \sat solver then confirms (in around 
$2$~s\footnote{On a computer running Linux, with an Intel i7-5500U 
CPU 2.40~GHz $\times$ 4 and 16GB.}) 
that there is no model (i.e., there is no counterexample), confirming 
that the reduction is correct for $k=20$, for \emph{any} graph with 
$50$ vertices. 
\end{example}

The example above confirms that, in certain cases, we can automatically 
justify the correctness of a reduction between two \np-complete problems
for bounded instance sizes. 
This approach is uniform: the user does not have to give any proof 
hints---they just have to provide the reduction. However, 
two important caveats must be noted. First, verifying correctness in 
this manner does not constitute a general proof. Instead, it is, 
fundamentally, an enumeration-based argument confirming the reduction's 
validity only for the given size of input instances. Second, for some 
reductions, using a relationship between the potential certificates 
for the two problems is still insufficient, as we will illustrate with 
the following, more elaborated examples.

\begin{example}[Verification of 3\sat to 3-colouring reduction] 
3\sat is a variant of \sat problem in which each clause has exactly 
three literals. 

For a graph $G$ with $n$ vertices there is a {\em $k$-colouring} if 
the graph's vertices can be labeled by $k$ colours such that no two adjacent 
vertices share the same colour. For $k>2$, the {\em $k$-colouring problem} 
is \np-complete. 

We can prove that the $3$-colouring problem is \np-complete by showing 
that it belongs to \np and by reducing 3\sat to it (assuming that it is 
already known that 3\sat is \np-complete, see Example \ref{ex:sat-3sat}). 
In the standard reduction, the resulting graph is constructed as follows 
\cite{AlgorithmDesignKleinberg}: 

\begin{itemize}
\item Three nodes\footnote{Sometimes called True, False, and Neutral or 
Base node.} $T$, $F$, $N$ are constructed and connected by edges. 
\item For each variable $v$ from the input 3\sat formula, the nodes 
$v$ and $\neg v$ are constructed, connected, and both connected to 
the node $N$. 
\item For each clause from the input 3\sat formula, one \say{gadget 
graph} (see figure below) is created, with its three \say{inputs} 
connected to three nodes in the clause, and with its output connected 
to both the node $N$ and the node $F$.
\end{itemize}

The input 3\sat formula is satisfiable iff the resulting graph is 
3-colourable. The figure below illustrates the reduction for the input 
3\sat formula 
$(u \vee \neg v \vee w) \wedge (v \vee x \vee y)$.\footnote{Example from
\url{https://cgi.csc.liv.ac.uk/~igor/COMP309/3CP.pdf}}

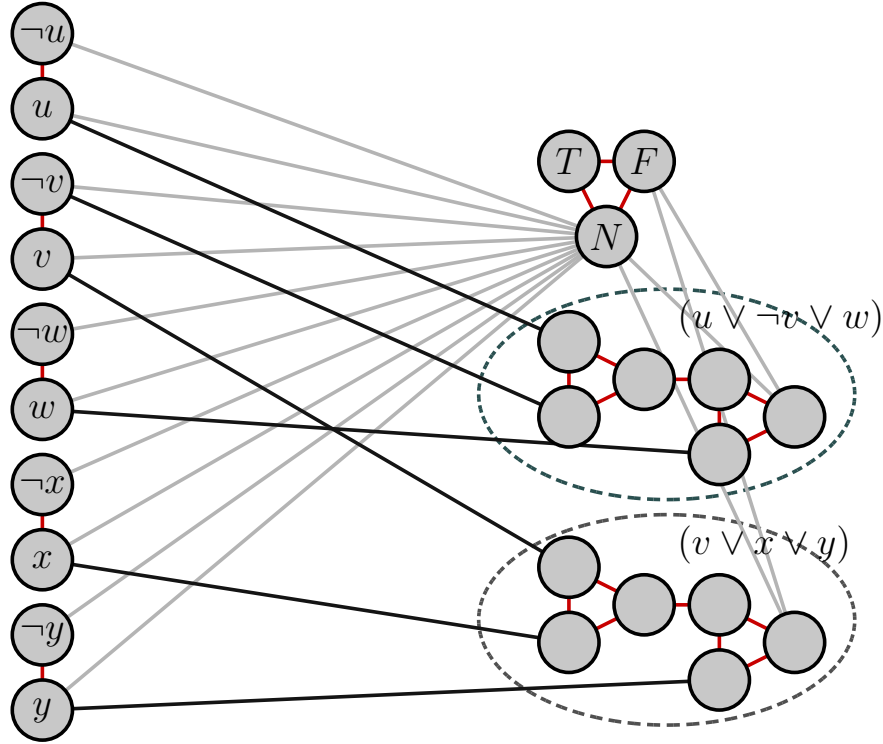
\begin{center}
\input{3SAT23coloring.tkz}
\end{center}

The following \ursa code implements the above reduction. The input clauses 
are assumed to be in the array \verb|nC|, but note that the code works with 
symbolic input formulae, not concrete ones. The values \verb|nC[i][0]|, 
\verb|nC[i][1]|, \verb|nC[i][2]| are indices of all three literals 
in the $i$-th 3\sat clause (from the sequence $x_0$, $\neg x_0$, $x_1$, 
$\neg x_1$, $\ldots$). The output graph adjacency matrix is also symbolic 
and stored in the array \verb|bE_colouring|. 

{\scriptsize
\begin{tcolorbox}[breakable,colback=instance,left=0mm,boxrule=0.1pt]
\begin{verbatim}
/* Specification of size of instance of 3SAT */
nN_3SAT = 5;
nClauses = 2;
\end{verbatim}
\end{tcolorbox}

\begin{tcolorbox}[breakable,colback=certification,left=0mm,boxrule=0.1pt]
\begin{verbatim}
/*** Certificate verification for 3SAT ***/
b3SAT = true;
for(ni=0; ni<nClauses; ni++) {
  b = false;
  for(nj=0; nj<3; nj++) {
    for(nk=0; nk<nN_3SAT; nk++) {
      b ||= ite(nC[ni][nj] == 2*nk,  bV[nk],  false);
      b ||= ite(nC[ni][nj] == 2*nk+1, !bV[nk],  false);
    }
  }
  b3SAT &&= b;
}

b3SATVarSet = true;
for(ni=0; ni<nClauses; ni++) {
  for(nj=0; nj<3; nj++) {
      b3SATVarSet &&= (nC[ni][nj] < 2*nN_3SAT);
  }
}
\end{verbatim}
\end{tcolorbox}

\begin{tcolorbox}[breakable,colback=reduction,left=0mm,boxrule=0.1pt]
\begin{verbatim}
/*** Reduction from 3SAT to 3-colouring ***/
/*
  Target graph:
  Nodes: 
     0: nNodeFalse, 1: nNodeTrue, 2: nNodeNeutral
  Nodes (for each variable - two graph nodes): 
     3: v0, 4: ~v0, ... 3 + 2*(n-1): v_{n-1}, 4 + 2*(n-1): ~v_{n-1} (for n==nN_3SAT))
  Gadget nodes (for each clause one gadget with 6 nodes):
     3+2*n, 4+2*n, 5+2*n, 6+2*n, 7+2*n, 8+2*n 
     ...
     3+2*n+6*(c-1), 4+2*n+6*(c-1), 5+2*n+6*(c-1), 6+2*n+6*(c-1), 7+2*n+6*(c-1), 8+2*n+6*(c-1) 
*/
nNodesColouring = 3 + 2*nN_3SAT + 6*nClauses;

nNodeFalse   = 0;
nNodeTrue    = 1;
nNodeNeutral = 2;

/* Initially reset all edges */
for(ni=0; ni < nNodesColouring; ni++) {
  for(nj=ni+1; nj < nNodesColouring; nj++) {
    bE_colouring[ni][nj] = false;
  }
}

/* Connect nodes T, N, F */
bE_colouring[nNodeFalse][nNodeTrue] = true;
bE_colouring[nNodeFalse][nNodeNeutral] = true;
bE_colouring[nNodeTrue][nNodeNeutral] = true;

/* Connect nodes v, ~v, N */
for(ni=3; ni<3 + 2*nN_3SAT; ni += 2) {
  bE_colouring[nNodeNeutral][ni] = true;
  bE_colouring[nNodeNeutral][ni+1] = true;
  bE_colouring[ni][ni+1] = true;
}

/* For each clause make a "gadget graph"
-0\
 | 3-4\ /N
-1/  | 5
    -2/ \F
*/
for(ni=0; ni<nClauses; ni++) {
   nS = 3 + 2*nN_3SAT + 6*ni;
   bE_colouring[nS][nS+1] = true;   
   bE_colouring[nS][nS+3] = true;
   bE_colouring[nS+1][nS+3] = true;
   bE_colouring[nS+3][nS+4] = true;
   bE_colouring[nS+2][nS+4] = true;
   bE_colouring[nS+2][nS+5] = true;  
   bE_colouring[nS+4][nS+5] = true;
   bE_colouring[nNodeFalse][nS+5] = true;       
   bE_colouring[nNodeNeutral][nS+5] = true;  
   for(nj=0; nj<2*nN_3SAT; nj++) {
     bE_colouring[3+nj][nS]   = (nC[ni][0] == nj);
     bE_colouring[3+nj][nS+1] = (nC[ni][1] == nj);
     bE_colouring[3+nj][nS+2] = (nC[ni][2] == nj);
   }
}

/*** Certificate reduction between 3SAT and 3-colouring ***/
bCertificateReduction = true;
for(ni=0; ni<nN_3SAT; ni++) {
  bCertificateReduction &&= !((nColour[3+2*ni]==nColour[nNodeTrue]) ^^ bV[ni]);
}
\end{verbatim}
\end{tcolorbox}

\begin{tcolorbox}[breakable,colback=certification,left=0mm,boxrule=0.1pt]
\begin{verbatim}
/*** Certificate verification for 3-colouring ***/
bNumberOfColours = true;
for(ni=0; ni<nNodesColouring; ni++) {
  bNumberOfColours &&= (nColour[ni] < 3);
}

b3colouring = true;
for(ni=0; ni<nNodesColouring; ni++) {
  for(nj=ni+1; nj<nNodesColouring; nj++) {
    b3colouring &&= ite(bE_colouring[ni][nj], nColour[ni] != nColour[nj], true);
  }
}
\end{verbatim}
\end{tcolorbox}

\begin{tcolorbox}[breakable,colback=certification,left=0mm,boxrule=0.1pt]
\begin{verbatim}
assert(b3SATVarSet && bNumberOfColours && !b3SAT && b3colouring && bCertificateReduction);
\end{verbatim}
\end{tcolorbox}
}

The final values of \verb|b3SATVarSet| and \verb|bNumberOfColours| are 
propositional formulae ensuring that certificates are legal (i.e., 
there are only $n$ propositional variables and that there are only $3$ 
colours). The final value of \verb|b3SAT| evaluates to a propositional formula 
such that \verb|b3SAT && b3SATVarSet| is satisfiable iff, for the given 
(symbolic) input 3\sat instance, the answer is \say{yes}. Conversely, the final value 
of \verb|b3colouring| is a propositional formula that is satisfiable iff, 
for the instance obtained by reduction, the answer is \say{yes}. Notice 
that in the above \ursa code, only the size of the input instance is 
provided. Consequently, the elements of the input 3\sat formula are 
treated as symbolic values, meaning that \verb|b3SAT| and \verb|b3colouring| 
collectively encompass all possible input formulae of the given size.
Following the previous examples, one may be tempted to assert

\noindent
\verb|assert(b3SATVarSet && bNumberOfColours && (b3SAT ^^ b3colouring)| \\
\verb|       && bCertificateReduction)|, 

\noindent
hoping that the resulting unsatisfiability (\say{no solution}) would 
ensure the correctness of the reduction. However, this is wrong: 
the formulas \verb|b3SAT| and \verb|b3colouring| are equisatisfiable
(similarly as in Example \ref{ex:clique_to_vertexcover_verification}), 
but in this case they do not share all variables. The variables occurring 
in \verb|b3colouring| but not in \verb|b3SAT| are not constrained by 
\verb|bCertificateReduction| and hence the above assertion does not 
guarantee correctness. Still, since the set of variables in 
\verb|b3colouring| is a strict superset of the set of variables in 
\verb|b3SAT|, if \verb|b3colouring| is true in some valuation, 
\verb|b3SAT| will be true too. Effectively, this means that it can be 
easily shown that the reduction is sound using the assertion:

\noindent
\verb|assert(b3SATVarSet && bNumberOfColours && !b3SAT && b3colouring| \\
\verb|       && bCertificateReduction)|.

\noindent
Soundness means that if the reduced (target) instance is a \say{yes} 
instance (e.g., satisfiable), then the original problem instance must 
also be a \say{yes} instance. This soundness property is vital because 
it ensures that the reduction does not introduce \say{false positives} 
(i.e., a solvable target instance from an unsolvable source instance). 
While this check ensures soundness only for a specific input size, 
it does so for {\em all} 3\sat formulae of that size.
By changing the parameters (number of variables and clauses), we can 
check the reduction soundness for any input size. For instance, for 
$10$ variables and $15$ clauses, \ursa generates a formula with $12792$ 
variables and $57926$ clauses. A \sat solver then confirms, in around 
42~s, that there is no model, confirming that the reduction is sound 
for this instance size, for \emph{any} 3\sat formula.
\end{example}

In the Appendix, an additional reduction is given: from \sat to 3\sat.

% ***************************************************************************
% ***************************************************************************
\section{Development Workflow}
\label{sec:Workflow}
% ***************************************************************************
% ***************************************************************************

The given examples demonstrate that reductions can be rather complex
(and modern, research-level reductions are even more so) 
and their implementation in \ursa can be challenging. How exactly is 
this implementation process carried out in practice? Specifying 
(partially, i.e., symbolically) the input is straightforward. Specifying 
the certificate verification procedure is usually also straightforward 
and, moreover, can be developed and tested independently of the reduction 
itself. The reduction component, however, is where implementation can become 
challenging. Developing the reduction resembles general programming: 
we write it, run/test it, fix any bugs, and repeat the cycle. A bug is 
exposed when the assertion confirming incorrectness (or unsoundness) 
succeeds and yields a model. Indeed, the key assertions, when they are not
met, imply correctness (or soundness) up to some input size. In other
words, the following holds:

\begin{center}
If the key assertion is met, then the reduction is not correct.
\end{center}

\noindent
Therefore, if a model (i.e., a countermodel for correctness or soundness) 
exists, we use it to locate the error in the code, and then we re-verify 
the results. This is similar to testing; however, it is much more powerful
since it covers all instances up to some size. 

The reductions presented above were implemented following this pattern. 
This iterative process can also serve for the reduction design process 
itself, particularly when the reduction structure is not known beforehand.

As noted, for some reductions, full correctness can be simply verified 
for all inputs with concrete input sizes, whereas in other cases, only 
soundness can be treated this way. The former case---full correctness---arises 
when all variables defining the first certificate occur in 
the second, and vice versa. The latter case---soundness only---arises 
when all variables defining the first certificate occur in the second, 
but not the opposite. Typically, in practice, all variables defining the 
first certificate play some role in defining the second certificate. 
Therefore, we can usually expect that at least soundness can be verified 
in a simple way for concrete input sizes.

% ***************************************************************************
% ***************************************************************************
\section{Related Work}
\label{sec:related}
% ***************************************************************************
% ***************************************************************************

There have been very few attempts to develop a comprehensive framework 
to aid in the development, analysis, or verification of reductions 
between \np-complete problems. Here we briefly describe them, and compare 
them to the approach presented in this paper.
 
For applications in education, an \say{online judge} is available to 
assess the correctness of student-submitted reductions between 
\np-complete problems \cite{CreusFG14}. This online system presents a 
set of \np-complete problem pairs for which the user must implement 
the reduction. To represent these reductions, a domain-specific, C-like, 
imperative programming language called REDNP was developed. The following 
are fragments of REDNP code that illustrate a reduction from the vertex cover 
problem to the dominating set problem:

\begin{center}
\begin{tcolorbox}[breakable,colback=reduction,left=0mm,boxrule=0.1pt]
{\scriptsize
\begin{verbatim}
in: struct {
  k: int
  nvertices: int
  edges: array of array [2] of int
}

out: struct {
  k: int
  edges: array of array [2] of string
  vertices: array of string
}

out.k = in.k;
foreach (edge; in.edges) {
  u = edge[0]; v = edge[1];
  out.edges.push = u, v;
  out.edges.push = u, "new{u}{v}";
  out.edges.push = v, "new{u}{v}";
}
\end{verbatim}
}
\end{tcolorbox}
\end{center}

\noindent
The system tests the given reduction by comparing two outputs for 
pre-defined test inputs obtained by solving them in the following ways: 
$(i)$ by the (predefined) reduction to \sat;
$(ii)$ by using the proposed reduction combined with the (predefined) 
reduction of the target problem to \sat. For each new \np-complete 
problem to be added to the system, there must be a reduction to \sat 
provided. Also, the test cases have to be hand-crafted or generated 
outside the framework. Although reductions to \sat are used in the 
testing context, a further step was not made: there is no proving of 
correctness of reductions, in general or for bounded input sizes. The 
system does not use or have certifiers (that could demonstrate that 
the considered problems belong to the class \np), even though they 
are a necessary component for showing \np-completeness. 
The complexity of reductions is implicit, and can be inferred from the 
REDNP code.

Karp is a declarative language for programming and testing \np 
reductions \cite{ZhangHD22}. Karp is a Racket-style framework that 
can be used to define computational problems and reductions between 
them. The following is a Karp 3\sat problem definition:

\begin{center}
\begin{tcolorbox}[breakable,colback=reduction,left=0mm,boxrule=0.1pt]
{\scriptsize
\verb|#lang karp/problem-definition| \\ 
\verb|(require karp/lib/cnf| \\
\verb|         karp/lib/mapping)|\\ 
\verb|(decision-problem #:name 3sat|\\
\verb|  #:instance ([|$\varphi$\verb| is-a (cnf #:arity 3)])|\\
\verb|  #:certificate (mapping|\\
\verb|                 #:from (variables-of| $\varphi$\verb|)|\\
\verb|                 #:to (the-set-of boolean)))|\\
\verb|; 3-\sat verifier definition|\\
\verb|(define-3sat-verifier a c^a|\\
\verb|  (|$\forall$\verb| [c |$\in$\verb| (clauses-of (|$\varphi$\verb| a))]|\\
\verb|    (|$\exists$\verb| [l |$\in$\verb| (literals-of c)]|\\
\verb|      (or|\\
\verb|         (and|\\
\verb|            (positive-literal? l)|\\
\verb|            (c^a (underlying-var l)))|\\
\verb|         (and|\\
\verb|            (negative-literal? l)|\\
\verb|            (not (c^a (underlying-var l))))))))|\\
}
\end{tcolorbox}
\end{center}

\noindent
The type system helps Karp programmers avoid simple mistakes when defining 
new problems. There is support for testing reductions in Karp through 
automatically generated instances. For deriving solvers, Karp utilizes 
the solver-aided programming system Rosette: Karp translates a problem 
definition into a Rosette function that, given a concrete problem instance, 
symbolically solves for a certificate. Certificates can be very complex 
(potentially including mappings, sets, and graphs), which is noteworthy 
since the types of Rosette's symbolic variables are limited to those 
supported by constraint solvers. Karp can automatically test the 
correctness of reductions using individual test instances, but it does 
not provide support for proving their correctness, in general or for
bounded input sizes. Karp cannot cover problems involving data types 
such as strings, real numbers, and functions.

In order to create a natural and expressive class of reductions, a graphical 
specification language, {\em cookbook reductions}, is introduced 
\cite{GrangeVVZ24}. This language is based on common building blocks 
used in computational reductions. Cookbook reductions are sufficiently 
expressive to cover a number of classical graph reductions, and they can 
be formally expressed as quantifier-free first-order interpretations. The 
verification of reductions for some classes of problems can be formulated 
as a satisfiability question for a decidable logic. This work is more of 
a theoretical nature: a prototype of the proposed approach is available,
but it supports only a limited number of cookbook reductions and 
problems, and automation of proving correctness is not supported.

We are not aware of any new \np-completeness results confirmed through 
the use of the above approaches.  

The \ursa-based approach presented here bears some resemblance to the 
above methodologies of REDNP, Karp, and cookbook reductions, but still 
there are significant differences. \ursa is a general-purpose constraint 
solving system, not a tool dedicated solely to reductions. 
Its underlying language is C-like and C code for checking certificates
and reductions can be used verbatim or almost verbatim. \ursa is 
distributed as an open-source, stand-alone desktop system that does 
not require dependencies on other software. Regarding the programming paradigm, 
unlike REDNP (imperative) and Karp (declarative), \ursa utilizes a 
specific imperative-declarative language structure. Regarding the interface, 
like REDNP and Karp, and in contrast to the graphical nature of cookbook 
reductions, \ursa is text-based. Similar to REDNP, but unlike Karp, 
\ursa does not offer automated testing for reductions. A key advantage 
is that \ursa allows the user to verify the correctness and/or soundness 
of the reductions for specified input sizes, a feature not supported in other 
approaches. Like the above systems, \ursa does not support strings or real numbers. 

With the highest possible level of trust, proofs of \np-completeness can be 
formalized within proof assistants such as Isabelle, Coq, and Lean. However, 
fully formalizing the entire scope---starting from the foundations of 
computation and computational complexity---remains extremely difficult 
\cite{RassJHH24, CoqGaherK21}.
Isabelle/HOL features a robust framework that tracks 
computational complexity and individual \np-complete problems. This includes 
formal definitions of deterministic multi-tape Turing machines, complexity 
classes, the Cook-Levin theorem, reductions, and \np-completeness proofs for 
several specific problems \cite{Cook_Levin-AFP}. Similar developments in 
complexity theory, \np-completeness, and the Cook-Levin theorem exist for the 
Coq interactive theorem prover \cite{CoqGaherK21}, alongside emerging 
computational complexity developments in Lean \cite{LeanSimas}. 
Additionally, some formalizations verify specific reductions by bypassing 
explicit complexity frameworks, omitting a formalized treatment of time 
complexity \cite{KreuzerN23, CoqKatz}. Notably, some of these formal 
achievements have revealed errors in earlier, pen-and-paper 
proofs \cite{KreuzerN23}. All of these developments are very recent. The 
total number of problems proved to be \np-complete within proof assistants 
remains low, and these results are generally confined to the interactive 
theorem proving community. The vast majority of newly published 
\np-completeness proofs are still traditional, informal arguments expressed 
in natural language. This disparity indicates that interactive theorem 
proving is still not easily accessible to most computer scientists and 
students working in complexity theory.

While other constraint programming systems could similarly be used 
to define and verify certificates and reductions (up to a certain 
input size), we are not aware of frameworks developed around other 
constraint solvers or specification languages nor of individual 
applications for justifying reductions between \np-complete problems.
Naturally, higher-level languages such as MiniZinc \cite{NethercoteSBBDT07} 
would be more suitable for this task than low-level languages such 
as SMT-LIB \cite{BarrettFT-SMTLIB}. 
However, \ursa may hold a comparative advantage since its underlying 
language is C-like, C code can be imported and exported almost verbatim,
and there is no need to learn some custom specification language.
Also, thanks to its imperative-declarative nature, the \ursa-based approach 
is suitable for ensuring the needed time complexity of certificate 
verification procedure and reduction procedures. Namely, it is 
generally easier to study time complexity in imperative languages 
because they provide explicit, step-by-step control, which aligns 
more directly with how time complexity is analysed through counting 
operations. 

There are other approaches and software systems related to reductions:
approaches for finding reductions algorithmically 
\cite{CrouchIM10,JordanK13}, approaches for reductions of \np-complete 
problems to STRIPS \cite{PorcoMB11}, educational support systems with 
modules for learning reductions in undergraduate courses 
\cite{Kneisel2024}, sets of \np-completeness reductions used in a study 
on using large language models in educational tools \cite{HerwigHKW25}, 
datasets composed of proofs of \np-completeness reductions for training 
and benchmarking the performance of large language models 
\cite{dicicco2025karpdataset}, etc.

% ***************************************************************************
% ***************************************************************************
\section{Conclusions and Future Work}
\label{sec:conclusions}
% ***************************************************************************
% ***************************************************************************

In this paper, we presented a methodology for designing, analyzing, and 
justifying reductions between \np-complete problems using \ursa, an 
open-source, imperative-declarative constraint-solving system based on 
\sat. Through several illustrative examples, we demonstrated 
the methodology and the following key features:

\begin{itemize}
\item The \ursa system is suitable for the simple creation of solvers 
for \np-complete problems, requiring only the certificate verification 
procedure.

\item The \ursa language is sufficiently expressive for representing 
reductions between \np-complete problems, including complex 
ones, in a rigorous way. Reductions can be checked for correctness 
using (hand-crafted) test instances.

\item The \ursa system can be used to automatically verify a reduction 
between two \np-complete problems for instances up to a certain 
size. For some problems, this justification ensures full correctness 
(both soundness and completeness), while for others, the justification 
ensures soundness only.
\end{itemize}

This methodology offers distinct advantages over alternative frameworks 
by providing straightforward, operational verification. Rather than testing 
isolated instances, it covers all inputs of a specified size. 
Because \ursa translates algorithmic constraints into propositional 
logic---requiring loop unrolling and fixed bit-vector bounds---this bounded 
verification is inherently non-uniform and does not constitute a general 
mathematical proof. 
However, this non-uniformity serves as a practical advantage: it enables 
rapid, automated checks that quickly expose logical errors in reduction 
directions or certificate mappings. Consequently, \ursa functions as an 
efficient prototyping environment that allows for the rapid debugging 
and validation of complex reductions, thereby building confidence in their 
correctness before developing a general informal or formal proof. 
Furthermore, because polynomial time complexity is essential for 
\np-completeness proofs, \ursa's restrictions on loop formation make it 
straightforward to confirm that both the verification and reduction 
procedures run in polynomial time.

Ultimately, the proposed workflow fills a crucial gap between informal
descriptions of reductions and fully formalized foundational proofs. 
It supplements human insight without attempting to replace it, serving 
as a valuable tool for researchers proving new \np-completeness results, 
as well as an educational resource for students exploring the relationships 
between these problems. 

While applying our approach to \np-completeness proofs in recent research 
publications is beyond the scope of this paper, it represents our primary 
target for future work. As a longer term goal, we consider implementing 
the presented approach within a proof assistant to reach a higher level 
of trust.

\section*{Acknowledgements}
We are grateful to Luka Markovi\'c for useful comments and ideas
regarding topics presented in this paper.

\bibliographystyle{plain}
%\bibliography{pj}

\newpage

\appendix

\section{Verification of \sat to 3\sat reduction}

\begin{example}[Verification of \sat to 3\sat reduction]
\label{ex:sat-3sat}
The 3\sat problem is in \np, and also \np-hard, which can be shown by 
reducing \sat to it (in polynomial time). The standard \sat to 3\sat 
reduction is as follows. Let $x$ be an instance of \sat. 
Each clause of $x$ with a length different from $3$ will be transformed 
into a set of clauses of length 3: 

\begin{itemize}
\item A clause with one literal $l_1$ is replaced by four clauses:
$l_1 \vee v \vee u$,
$l_1 \vee v \vee \neg u$,
$l_1 \vee \neg v \vee u$, and
$l_1 \vee \neg v \vee \neg u$, where $v$ and $u$ are new variables.

\item A clause with two literals $l_1 \vee l_2$ is replaced by two 
clauses: $l_1 \vee l_2 \vee v$ and $l_1 \vee l_2 \vee \neg v$, where $v$
is a new variable.

\item A clause with more than three literals $l_1 \vee l_2 \vee l_3 \vee \ldots \vee l_k$ 
is replaced by two clauses: 
$l_1 \vee l_2 \vee v$ and 
$l_3 \vee \ldots \vee l_k \vee \neg v$, where $v$ is a new variable, and the 
latter clause is further transformed and eventually replaced by clauses of 
length 3.
\end{itemize}

\noindent
As said above, one \sat clause yields 1, 2, or 4 clauses, while if the 
starting clause has more than 3 literals, the resulting clauses will be 
recursively transformed further until they are 3\sat. If $t(k)$ denotes 
the total number of clauses eventually produced by a clause of length 
$k$, the following relationship holds:

$$t(k) =
\left\{
\begin{array}{ll}
4, & k=1 \\
2, & k=2 \\
1, & k=3 \\
k-2, & k > 3
\end{array}
\right.
$$

\noindent
If there are $n$ variables and $n>1$, then the longest possible \sat 
clause contains all of them and all of their negations, so its length 
is $2n$ and it produces $2n-2$ 3\sat clauses.

In the above way, an instance of \sat is transformed into an instance 
of 3\sat. The reduction is polynomial and the former is satisfiable 
iff the latter is satisfiable, hence 3\sat is \np-complete.

The following \ursa code implements the above reduction from \sat to 
3\sat. The input \sat clauses are assumed to be in the array \verb|bC_SAT|, 
while the output 3\sat clauses are stored in the array \verb|nC_3SAT|. The 
chosen representations for \sat and 3\sat instances are significantly 
different: \verb|bC_SAT[i][j]| is Boolean, and true if the $i$-th clause 
contains the $j$th literal from the sequence: $x_0$, $\neg x_0$, $x_1$, 
$\neg x_1$, $\ldots$, while \verb|nC_3SAT[i][0]|, \verb|nC_3SAT[i][1]|, 
\verb|nC_3SAT[i][2]| are indices of all three literals in the $i$-th 
3\sat clause. Note that the given code works with either symbolic or 
concrete values. If only the size of the input instance is given, the 
input array encompasses all possible inputs of that size. This and the 
fact that in \ursa if-statements can be applied only on ground expressions, 
and arrays can be accessed only with ground indices, led to some specifics 
in the implementation. If the input clause has at most three literals, it 
leads to at most four clauses in the output array. Since the processing 
is symbolic, there are always four slots for output clauses. If the input 
clause is of the form $l_1 \vee l_2 \vee l_3 \vee \ldots \vee l_k$, it 
produces the output clause $l_1 \vee l_2 \vee v$ (plus three tautological 
clauses), and is replaced by $l_3 \vee \ldots \vee l_k \vee \neg v$, as
illustrated by the following table:

\begin{center}
\begin{tabular}{|c|c|c|} \hline
input: \verb|bC_SAT| array & & output: \verb|nC_3SAT| array \\ \hline
$l_1 \vee l_2 \vee l_3 \vee \ldots \vee l_k$ & & \\ \hline
                                             & & \\
                                             & $\downarrow$ & \\ \hline   
$l_3 \vee \ldots \vee l_k \vee \neg v$       & & $l_1 \vee l_2 \vee v$ \\ \hline   
                                             & & $p \vee \neg p \vee \neg p$ \\ \hline    
                                             & & $p \vee \neg p \vee \neg p$ \\ \hline   
                                             & & $p \vee \neg p \vee \neg p$ \\ \hline
\end{tabular}
\end{center}

\noindent
There could be $2n-2$ such iterations (if $n$ is the number of input 
variables). Again, since the execution is symbolic, if there are $C$ input 
clauses, the above code always produces $C \cdot (2n-2) \cdot 4$ output 
clauses (while some of them are trivial tautologies). The translation 
could be defined in some other way to save more space, but then it would
lead to more complicated code.

{\scriptsize
\begin{tcolorbox}[breakable,colback=instance,left=0mm,boxrule=0.1pt]
\begin{verbatim}
/*** Specification of size of instance of SAT problem ***/
nN_SAT = 10;
nClauses_SAT = 10;
\end{verbatim}
\end{tcolorbox}

\begin{tcolorbox}[breakable,colback=certification,left=0mm,boxrule=0.1pt]
\begin{verbatim}
/*** Certificate verification for SAT formula ***/
bSAT = true;
for(ni=0; ni<nClauses_SAT; ni++) {
  b = false;
  for(nj=0; nj<nN_SAT; nj++) {
    b ||= ite(bC_SAT[ni][2*nj],   bV_SAT[nj],  false);
    b ||= ite(bC_SAT[ni][2*nj+1], !bV_SAT[nj], false);
  }
  bSAT &&= b;
}

/*** Non-empty formula condition ***/
bNonEmpty_SAT = true;
for(ni=0; ni<nClauses_SAT; ni++) {
  b = false;
  for(nj=0; nj<2*nN_SAT; nj++) {
    b ||= bC_SAT[ni][nj];
  }
  bNonEmpty_SAT &&= b;
}
\end{verbatim}
\end{tcolorbox}

\begin{tcolorbox}[breakable,colback=reduction,left=0mm,boxrule=0.1pt]
\begin{verbatim}
/*** Reduction from SAT to 3SAT ***/

/* At most 2*nN_SAT-2 output SAT clauses can be obtained from one clause with > 3 literals */
/* At most 4 output 3SAT clauses can be obtained from one clause with <= 3 literals        */
nClauses_3SAT = nClauses_SAT * ite(nN_SAT > 1, 2*nN_SAT - 2, 4) * 4; 
nN_3SAT       = nN_SAT + nClauses_SAT * (2*nN_SAT - 3) * 2;

for(ni=0; ni<nClauses_SAT; ni++) { /* Clear-up all additional variables */
  for(nj=2*nN_SAT; nj<2*nN_3SAT; nj++) {  
    bC_SAT[ni][nj] = false;
  }
}

for(ni=0; ni<nClauses_3SAT; ni++) { /* Initialize all output clauses to simple tautologies */
  nC_3SAT[ni][0] = 0;
  nC_3SAT[ni][1] = 1;          
  nC_3SAT[ni][2] = 1;
}

for(ni=0; ni < nClauses_SAT; ni++) {
  for(nl=0; nl < 2*nN_SAT-2; nl++) { 
    nCount = 0;
    for(nk=0; nk < 2*nN_3SAT; nk++) {  /* Count literals in the clause */
      nCount += ite(bC_SAT[ni][nk], 1, 0);
    }
    nNewVar = nN_SAT + ni*(2*nN_SAT-2)*2 + nl*2;
    nOutC   = ni*(2*nN_SAT-2)*4 + nl*4;

    /* Handle clause with 1 literal */
    for(nj=0; nj<2*nN_3SAT; nj++) {
      nC_3SAT[nOutC+0][0] = ite(nCount == 1 && bC_SAT[ni][nj], nj, nC_3SAT[nOutC+0][0]);
      nC_3SAT[nOutC+1][0] = ite(nCount == 1 && bC_SAT[ni][nj], nj, nC_3SAT[nOutC+1][0]);
      nC_3SAT[nOutC+2][0] = ite(nCount == 1 && bC_SAT[ni][nj], nj, nC_3SAT[nOutC+2][0]);
      nC_3SAT[nOutC+3][0] = ite(nCount == 1 && bC_SAT[ni][nj], nj, nC_3SAT[nOutC+3][0]);
    }
    nC_3SAT[nOutC+0][1] = ite(nCount == 1, 2*nNewVar,   nC_3SAT[nOutC+0][1]);
    nC_3SAT[nOutC+0][2] = ite(nCount == 1, 2*nNewVar+2, nC_3SAT[nOutC+0][2]); 
    nC_3SAT[nOutC+1][1] = ite(nCount == 1, 2*nNewVar+1, nC_3SAT[nOutC+1][1]);
    nC_3SAT[nOutC+1][2] = ite(nCount == 1, 2*nNewVar+2, nC_3SAT[nOutC+1][2]);
    nC_3SAT[nOutC+2][1] = ite(nCount == 1, 2*nNewVar,   nC_3SAT[nOutC+2][1]);
    nC_3SAT[nOutC+2][2] = ite(nCount == 1, 2*nNewVar+3, nC_3SAT[nOutC+2][2]);
    nC_3SAT[nOutC+3][1] = ite(nCount == 1, 2*nNewVar+1, nC_3SAT[nOutC+3][1]);
    nC_3SAT[nOutC+3][2] = ite(nCount == 1, 2*nNewVar+3, nC_3SAT[nOutC+3][2]);

    /* Handle clause with 2 literals */
    nCount2 = 0;
    for(nj=0; nj<2*nN_3SAT; nj++) {
      nCount2 += ite(bC_SAT[ni][nj], 1, 0);
      nC_3SAT[nOutC+0][0] = ite(nCount==2 && nCount2==1 && bC_SAT[ni][nj], nj, nC_3SAT[nOutC+0][0]);
      nC_3SAT[nOutC+0][1] = ite(nCount==2 && nCount2==2 && bC_SAT[ni][nj], nj, nC_3SAT[nOutC+0][1]);
      nC_3SAT[nOutC+1][0] = ite(nCount==2 && nCount2==1 && bC_SAT[ni][nj], nj, nC_3SAT[nOutC+1][0]);
      nC_3SAT[nOutC+1][1] = ite(nCount==2 && nCount2==2 && bC_SAT[ni][nj], nj, nC_3SAT[nOutC+1][1]);    
    }
    nC_3SAT[nOutC+0][2] = ite(nCount == 2, 2*nNewVar,   nC_3SAT[nOutC+0][2]);
    nC_3SAT[nOutC+1][2] = ite(nCount == 2, 2*nNewVar+1, nC_3SAT[nOutC+1][2]);          

    /* Handle clauses with => 3 literals */
    nCount2 = 0;
    for(nj=0; nj<2*nN_3SAT; nj++) {
      nCount2 += ite(bC_SAT[ni][nj], 1, 0);
      nC_3SAT[nOutC][0] = ite(nCount >= 3 && nCount2 == 1 && bC_SAT[ni][nj], nj, nC_3SAT[nOutC][0]);
      nC_3SAT[nOutC][1] = ite(nCount >= 3 && nCount2 == 2 && bC_SAT[ni][nj], nj, nC_3SAT[nOutC][1]);
      nC_3SAT[nOutC][2] = ite(nCount == 3 && nCount2 == 3 && bC_SAT[ni][nj], nj, nC_3SAT[nOutC][2]);        
    }
    nC_3SAT[nOutC][2] = ite(nCount > 3, 2*nNewVar, nC_3SAT[nOutC][2]);
    nCount2 = 0;
    for(nj=0; nj<2*nN_3SAT; nj++) {
      nCount2 += ite(bC_SAT[ni][nj], 1, 0);    
      bC_SAT[ni][nj] = ite(nCount > 3 && (nCount2 == 1 || nCount2 == 2) && bC_SAT[ni][nj], 
                           false, 
                           bC_SAT[ni][nj]);
    }
    bC_SAT[ni][2*nNewVar+1]   = ite(nCount > 3, true, bC_SAT[ni][2*nNewVar+1]);  
	
    /* Clear fully translated clauses */ 
    for(nj=0; nj<2*nN_3SAT; nj++) {  
      bC_SAT[ni][nj] = ite(nCount <= 3, false, bC_SAT[ni][nj]);
    }
  }
} 

/*** Certificate reduction from SAT to 3SAT ***/
bCertificateReduction = true;
for(ni=0; ni<nN_SAT; ni++) {
  bCertificateReduction &&= !(bV_SAT[ni] ^^ bV_3SAT[ni]);
}
\end{verbatim}
\end{tcolorbox}

\begin{tcolorbox}[breakable,colback=certification,left=0mm,boxrule=0.1pt]
\begin{verbatim}
/*** Certificate verification for 3SAT formula ***/
b3SAT = true;
for(ni=0; ni<nClauses_3SAT; ni++) {
  b = false;
  for(nj=0; nj<3; nj++) {
    for(nk=0; nk<nN_3SAT; nk++) {
      b ||= ite(nC_3SAT[ni][nj] == 2*nk,  bV_3SAT[nk],  false);
      b ||= ite(nC_3SAT[ni][nj] == 2*nk+1, !bV_3SAT[nk],  false);
    }
  }
  b3SAT &&= b;
}

/*** Legal 3SAT formula ***/
b3SAT_legal = true;
for(ni=0; ni<nClauses_3SAT; ni++) {
  for(nj=0; nj<3; nj++) {
     b3SAT_legal &&= (nC_3SAT[ni][nj] < 2*nN_3SAT);
  }
}
\end{verbatim}
\end{tcolorbox}

\begin{tcolorbox}[breakable,colback=certification,left=0mm,boxrule=0.1pt]
\begin{verbatim}
/*** Reduction verification ***/
assert(bNonEmpty_SAT && b3SAT_legal && (!bSAT && b3SAT) && bCertificateReduction); 
\end{verbatim}
\end{tcolorbox}
}

\noindent
Commands of the form \verb|n = ite(b, na, n);| are frequently used; they 
are necessary (instead of the more natural \verb|if (b) n=na;|) because, 
in \ursa, \verb|if|-state\-ments can be applied only on ground expressions.

The final values of \verb|bNonEmpty_SAT| and \verb|b3SAT_legal| are 
propositional formulae that ensure that certificates are legal (i.e., 
there are only $n$ propositional variables involved). The final value 
of \verb|bSAT| is a propositional formula that is satisfiable iff the 
(symbolic) input formula is satisfiable. Conversely, the final value of 
\verb|b3SAT| is a propositional formula that is satisfiable iff the 
3\sat formula obtained by the reduction is satisfiable. As mentioned, 
only the size of the input instance is provided, so both expressions 
are treated as symbolic values and thus encompass all possible input 
formulae of the given size. Following lessons from the previous two 
examples, these formulae are equisatisfiable, and the latter is dependent 
on some variable not occurring in the former. Because of that, we can 
simply verify only the soundness of the reduction, as shown in the 
given assertion. For instance, for all formulae with 10 clauses and 
10 variables, the soundness is verified in 450~s (by checking a \sat 
formula with 1309972 variables and 10470500 clauses). A brute force 
enumeration testing (of all possible inputs), of course, could not 
come anywhere near this performance.
\end{example}

\end{document}